\documentclass[a4paper,11pt]{article}
\pdfoutput=1

\usepackage{jinstpub}     % two modifications in .sty by PF to eliminate compile errors

\usepackage{graphicx}
\usepackage{xspace}
\usepackage{booktabs}     % for nice tableslarg
\usepackage{tikz-timing}  % for timing diagram
\usepackage{enumitem}     % for \setlist

\usepackage[sort&compress,numbers]{natbib}

\newcommand\inquotes[1]{\lq#1\rq}
\newcommand\Fig     [1]{Figure~\ref{#1}\xspace}

\newcommand\fig     [1]{Fig.~\ref{#1}}
\newcommand\tab     [1]{table~\ref{#1}}

\def\nm{\mathrm{\,   nm}}
\def\um{\mathrm{\,\upmu m}}
\def\mm{\mathrm{\,   mm}}
\def\cm{\mathrm{\,   cm}}

\def\ns{\mathrm{\,ns}}

\def\ms{\mathrm{\,ms}}
\def\s {\mathrm{\,s}}

\def \V  {\mathrm{\,    V}}

\def\Hz {\mathrm{\, Hz}}

\def\MHz{\mathrm{\,MHz}}

\def\khits {\,\mathrm{khits}}
\def\Mhits {\,\mathrm{Mhits}}

\def\degC  {\,^{\circ}\mathrm{C}}

\def\logicone{\texttt{high}\xspace}
\def\logiczero{\texttt{low}\xspace}

\newcommand {\FigureStandard}[3] {
 \begin{figure}[!t]
 \centering
 \includegraphics[width=0.95\linewidth]{figures/#1}
 \vskip-2mm
 \caption{#3} \label{#2}
 \end{figure}
}

\newcommand {\FigureVariable}[4] {
 \begin{figure}[!t]
 \centering
 \includegraphics[width=#4\linewidth]{figures/#1}
 \vskip-2mm
 \caption{#3} \label{#2}
 \end{figure}
}

\setlist[itemize]{
  parsep=1pt,   % vertical space between paragraphs in ONE item
  itemsep=0pt,  % ADDITIONAL space (to parsep) between items
  label=$\bullet$,
  topsep=1pt,
  labelindent=1mm, labelsep=2mm, leftmargin=*
}

\def\GND    {\texttt{GND}\xspace}
\def\VDD    {\texttt{VDD}\xspace}
\def\BIAS   {\texttt{BIAS}\xspace}

\def\CMD    {\texttt{CMD}\xspace}
\def\CLK    {\texttt{CLK}\xspace}
\def\SERIN  {\texttt{SERIN}\xspace}
\def\SEROUT {\texttt{SEROUT}\xspace}

\def\SOA    {\texttt{SEROUT1}\xspace}  % for timing diagram
\def\SOB    {\texttt{SEROUT2}\xspace}

\def\HITCOL {$\overline{\texttt{HitCol}}$\xspace}
\def\HITROW {$\overline{\texttt{HitRow}}$\xspace}
\def\SENDROW{\texttt{SendRow}\xspace}
\def\RESET  {\texttt{reset}\xspace}
\def\INJECT {\texttt{inject}\xspace}

\def\ffChipA   {72.5}
\def\ffModuleA {69}  % 68.9

\def\ffPlaneB  {71}  % 70.9

\def\ChipX {8046}
\def\ChipY {9032}

\title{Design of a Large Area Digital SiPM with High Fill Factor and Fully Serial Digital Readout for Single Photon Detection in Liquid Noble Gas Detectors}

\author{Peter Fischer,}
\author{Michael Keller}
\author{and Michael Ritzert}
\affiliation{ZITI, Institute of Computer Engineering, Heidelberg University, Im Neuenheimer Feld 368, 69120 Heidelberg, Germany}

\emailAdd{peter.fischer@ziti.uni-heidelberg.de}
 
\begin{document}

\abstract {
  We present a \inquotes{digital SiPM} photo-detection chip combining single photon sensitive avalanche photo diodes and CMOS readout electronics on a single die. The chip has a size of $\ChipX\times\ChipY\um^2$ with $\ffChipA\%$ of photo sensitive area. It is subdivided into $32\times30$ pixels with an average size of $250\times291\um^2$. For each photon hit, the chip records the pixel coordinate and the arrival time with a granularity of $\approx 10\ns$.
  Readout and chip control are purely digital, requiring only 4 CMOS signals. Several chips can be daisy chained and grouped on larger modules so that detector planes with $\approx70\%$ photo sensitive area can be build. Our chip may be used in experiments that need to detect rare scintillation events, for instance dark matter searches using liquid noble gases.   
}

\maketitle
\flushbottom


%===========================================================================
\section{Motivation}
%===========================================================================

The detection of small numbers of optical photons is an omnipresent challenge in many scientific experiments, for instance in liquid noble gas detectors, which are used for dark matter search \cite{DM1,DM2} or neutrino physics \cite{Dune}. In these applications, large tanks are filled with several cubic meters of liquid Xenon (at $-108\degC$) or Argon (at $-186\degC$) where the particles of interest interact and produce, among others, some visible photons in the short wavelength range ($178\nm$ (LXe)\cite{RefLXe}, $128\nm$ (Ar)\cite{RefLAr}). Due to the low photon number, highly sensitive  detectors and a large sensitive area (high fill factor) are required. The Dark Count Rate (DCR) of the photo detectors must be very low ($\le 0.01\Hz/\mathrm{mm}^2$) in order to limit the number of coincidences of noise hits, which may fake the signature of the very rare events of interest.

Existing experiments \cite{Experiment_XenT,Experiment_LZ,Experiment_Darkside20k} use specialized vacuum photo multiplier tubes (PMT) with excellent performance figures. These devices, however, can have lifetime issues and they contain radioactive isotopes which produce undesired background events \cite{PMT}.

Silicon-based Single Photon Avalanche Diodes (SPADs), which amplify the photon-gener\-ated electron-hole pair in an avalanche process, may be an alternative offering very low intrinsic radioactivity and longevity. Recent advances in the manufacturing technology improved the quantum efficiency in the UV range and the DCR at cold temperatures significantly \cite{FBK17,VUV1}. The most common devices are arrays of SPADs of $(10-100\um)^2$ area, each biased by a separate resistor, such that each SPAD can be operated at high gain in 'Geiger Mode'. These Silicon Photo Multipliers (SiPMs) (or Multi Photon Pixel Counters, MPPCs) have overall sizes of $(1-6\mm)^2$ containing many 1000s of individual SPADs.
Because the large voltage signal generated in the avalanche discharge of one SPAD is 'shared' between the many other SPADs, the (fast) signal component seen at the two device terminals is small. A fast, low noise amplifier is therefore required to maintain single photon sensitivity \cite{ColdAmplifier1,ColdAmplifier2,ColdAmplifier3}. The cooling of the relatively power-hungy amplifiers, the space required for the electronics, issues of material purity and the need for a large number of cables are challenges for system design. The further processing of the amplified analogue signals requires a large number of fast ADCs.

We propose to get rid of amplifiers and ADCs by integrating a digital readout directly onto the photo sensor. This can be achieved by merging SPADs and CMOS electronics on a \inquotes{digital SiPM} (DSiPM) or \inquotes{CMOS SPAD Array}. Several chip manufacturers offer such a device combination. We have chosen the technology offered by the Fraunhofer institute IMS in Duisburg / Germany because of its low dark count rate, which is of utmost importance for the target application. The IMS technology offers MOS devices with a minimum gate length of $350\nm$ and provides 4 metal layers.

An obvious challenge of the DSiPM approach is the reduction of the photo sensitive area due to the electronic circuitry. One goal of our design was therefore to maximized the fill factor by a highly optimized architecture, keeping the readout circuit as simple and small as possible. Our readout approach addresses two very different signal signatures occurring in liquid noble gas experiments:
\begin{itemize}
\item {\em Direct photons} from interesting physics events (often called the \inquotes{S1     
      signal}) or from background interactions occur randomly at a very low rate. The readout
      of these photons and of hits generated by dark noise can be simple and a moderate data throughput is sufficient. A precise time stamping of each hit is required for the coincidence analysis.
\item The {\em charges} also produced in some interactions are often drifted
      to a liquid-gas interface with an amplification region where they produce up to $\approx 10000$ \cite{DM2} {\em simultaneous} photons (the \inquotes{S2 signal}. These bursts are very rare so that their readout can take long. Time stamping does not need to be perfect for these hits because they can be clearly associated to one event.
\end{itemize}

The time resolution for the photon hits can be moderate ($10\ns$), because the arrival times are only used to make coincidences with a rather large coincidence time window due to the long photon flight times in the large experiments and because of multiple scattering in the liquid. A spatial resolution of $\lesssim 1\cm$ is sufficient.

The following chapter \ref{sect:chip} describes in detail the geometry and architecture of the chip which has been submitted for fabrication in late 2023. Some very first measurement are presented in chapter \ref{sect:measurements}. Chapter \ref{sect:discussion} compares the key properties to the requirements of dark matter search experiments like DARWIN and gives an outlook to possible module designs.

%===========================================================================
\section{Chip Description}\label{sect:chip}
%===========================================================================

The main goals of the chip development were
\begin{itemize}
\item A large fill factor, i.e.\ a maximized photosensitive area.
\item A low dark count rate (DCR) at the operating temperature.
      This mainly requires a good manufacturing technology. 
      In addition, it should be possible to turn off particularly noisy SPADs by lowering
      their bias voltage.
\item A continuous readout of the x/y position and the time of individual photon hits
      occurring at a low rate.
\item The ability to record and read out a bursts of many simultaneous photons, as they occur
      in the S2 signal.
\item A purely digital interface to ease the construction of larger modules with
	  multiple chips.
\item A minimal power dissipation to minimize the cooling effort in the experiment.
\end{itemize}

Fill factor can obviously be increased with a large chip, because many global elements like bonding pads, chip configuration, and readout circuitry are better shared. While the technology allows a maximum reticle size of $\approx 19\times19\mm^2$, our particular run was technically limited in height to $\approx 9\mm$ and we had to share the width with other designs. This limited the size of this prototype chip to $\approx 8\times9\mm^2$, see \fig{fig:fullchip}.

\FigureVariable{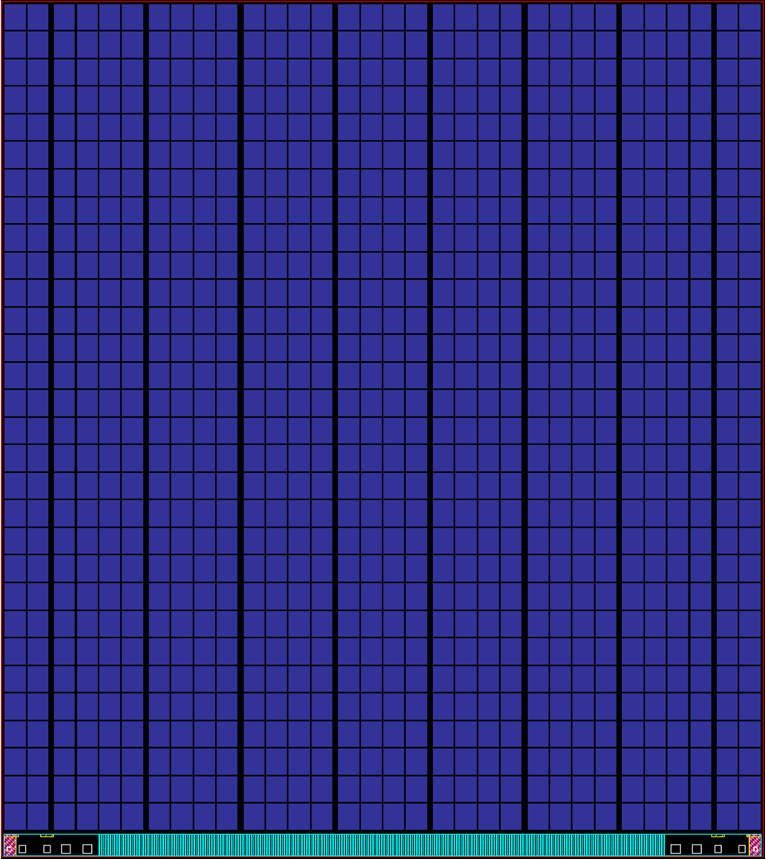}{fig:fullchip}{The submitted chip has $32\times30$ pixels in the active part with an area of $8000\times8730\um^2$.
Global circuitry and 10 wire bond pads are combined at the bottom in a strip of only $279\um$ height. With edges of $23\um$ between circuitry / SPADs and the dicing line, the total nominal chip size is $\ChipX\times\ChipY\um^2$.}{0.5}

%---------------------------------------------------------------------------
\subsection{SPAD and Pixel Size}
%---------------------------------------------------------------------------

An important design decision is the size of the individual SPADs. There are a number of arguments for small or large SPAD area:

\begin{itemize}
\item Large SPADs obviously lead to a higher fill factor because less signals must be processed,
      requiring less electronic circuitry.
\item Large SPADs degrade the position resolution, but that is in any case much 
      better than required in our primary target application.
\item Large SPADs increase the probability that multiple photons hit the {\em same} 
      SPAD during a measurement time window. The resulting SPAD discharge is counted as {\em one} hit, so that the additional photons remain undetected.
      This 'saturation effect' is one reason why standard SiPMs often have small SPADs.
      In our application, photon rates are so small that saturation may only occur for very large S2 signals.
\item Large SPADs operate at a higher gain due to their larger capacitance, developing more
      charge in the avalanches. This increases the chances for crosstalk and afterpulsing.
      Because of the very low occupancy in this application, and the possibility to read 
      out the individual SPADs, these issues can probably be handled very well by appropriate
      data analysis. 
\item When a SPAD contains a defect leading to a high dark count rate, this SPAD must be
      switched off electronically. 
      When the disabled SPAD is large, a significant active area is lost and fill factor suffers. For small SPADs, on the other hand, the area loss when turning off a SPAD is small, but the fill factor is intrinsically worse, because of edge losses and more electronics.
\end{itemize}

\begin{figure}[!t]
 \centering
 \includegraphics[width=0.52\linewidth]{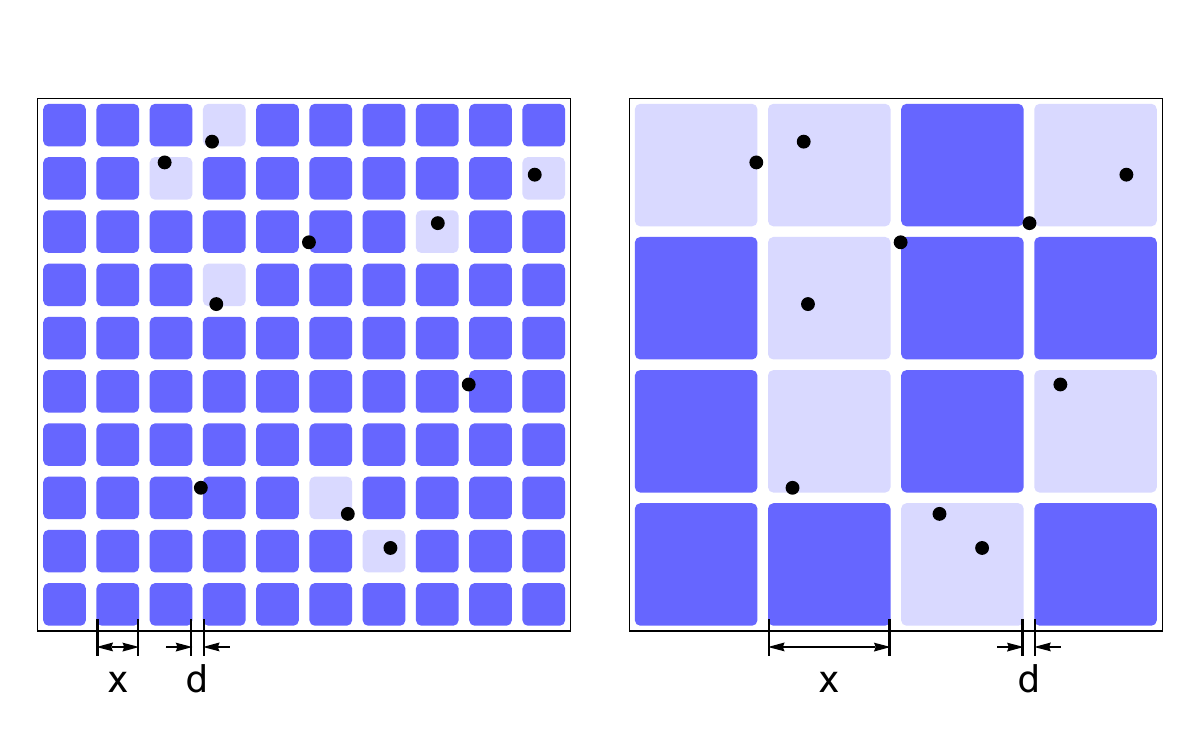}
 \hfill
 \includegraphics[width=0.46\linewidth]{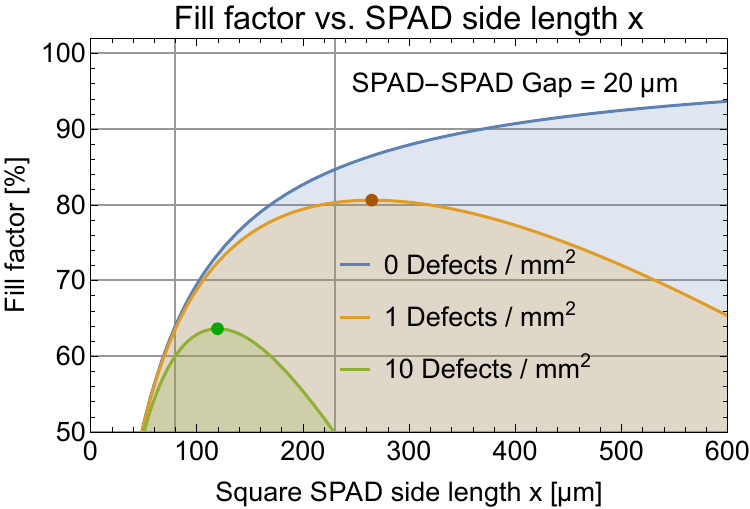}
 \vskip-2mm
 \caption{Left: Illustration of an area of $1\times 1\mm^2$ with SPADs of $80\um$ and $230\um$, respectively, with (exaggerated) gaps between SPADs of $20\um$ and $10$ random defects per $\mm^2$. SPADs containing a defect must be disabled and their active area is lost. Right: Fill factor as a function of SPAD side length for different defect densities according to eq.\,\ref{eq:ff} illustrating a clear optimum for a particular defect density. The fill factor of the smaller SPADs is worse in a defect free situation (blue) due to the gaps between SPADs, but becomes favorable in the presence of defects (green). } 
 \label{fig:optimization}
\end{figure}

The driving argument for our design was a maximum fill factor {\em after all noisy SPADs have been disabled}. As illustrated in \fig{fig:optimization}, the optimum SPAD size  depends on the density of 'fatal' defects and on other design parameters like the minimum SPAD-SPAD distance (and the area required for in-pixel and global electronic circuitry, not considered here). 
Assuming a defect density of $\epsilon$ (defects per $\mm^2$), the average number of defects on a SPAD of area $A$ is $A\epsilon$. The probability that such a SPAD is good, i.e.\ contains $0$ defects, is the Poisson value $P_{A\epsilon}(0) = e^{-A\epsilon}$. We assume square SPADs of side length $x$ with area $A=x^2$ and gaps between SPADs of $d$ for this illustration. From $N$ SPADs on a chip, only a fraction of $N e^{-A\epsilon}$ is usable on average, so that the active area is $A\,Ne^{-A\epsilon}$. This has to be compared to the total area of $N (x+d)^2$ so that the fill factor is
\begin{equation}
 \mathrm{FF}_{d,\epsilon}(x) = \frac{x^2}{(d+x)^2} e^{-\epsilon x^2}.\label{eq:ff}
\end{equation}
This function is plotted in \fig{fig:optimization} (right) for $d=20\um$ and three different defect densities $\epsilon$. For no defects (blue), SPADs should obviously be as large as possible. For higher defect densities, small SPADs are more robust.

The situation in a real chip is more complicated due to the presence of electronics per SPAD (the area of which is not well known a priori), of global circuits and bond pads. Calculations of the type shown above have been used as a starting point and the design has been iterated, taking into account other constraints like a height sufficient to fit all NMOS/PMOS in one column. The defect density $\epsilon$ has been obtained from measurements on several prototype chips, where it has been observed that $\epsilon$ is temperature dependent: Some defects with intermediate activation energies 'freeze out' at low temperatures and some noisy SPADs become usable when operated cold. The defect density at liquid Xenon temperature has been used to guide the design.

\FigureStandard{Pixel4.png}{fig:pixelgroup}{Group of 4 Pixels with shared readout circuit in the middle. The upper right quadrant of the readout electronics, highlighted in pink, is shown in detail in \fig{fig:pixellayout}.}

The position resolution delivered by the \inquotes{optimal} SPAD size is much better than required and the readout of many individual small SPADs would consume an unnecessary amount of circuity. We have therefore decided to group 9 SPADs (but with individual disable feature!) into one logical pixel by OR-ing their hit information. This number could be increased, but the reduction in circuitry is small. Note that this solution is different from a 9 times larger SPAD, because each of the 9 SPADs can be disabled {\em individually}, so that only 1/9th of the area is lost by a defect. (When such a pixel with multiple SPADs is measured to be noisy, the guilty SPADs can be found by enabling only one of the 9 SPADs at a time.)

In order to further reduce the space required for electronics, the logic circuit for 4 pixels has been merged together. This is beneficial because the spacing between SPADs can be smaller than the spacing between SPAD and transistors, due to the high voltage potential of the SPAD N-wells. In addition, control signals can be shared. This group of 4 pixels, each subdivided into 9 SPADs of nearly identical size, is shown in \fig{fig:pixelgroup}. Given the chip area constraints mentioned above, and the space needed for some global circuitry at the bottom, we could fit $8\times30$ such groups and thus $32\times30$ pixels on the chip.

The layout of the circuit between the SPADs has been highly optimized to fit in one vertical column of PMOS and two columns of NMOS, see \fig{fig:pixellayout}.\,A layout with hexagonal pixels for a theoretically still slightly higher fill factor has been tried, but has finally been rejected because of too many different SPAD shapes at the chip edges and because of the effort (and design rule violations) when routing with non-45 degree traces between the SPADs. 

\FigureVariable{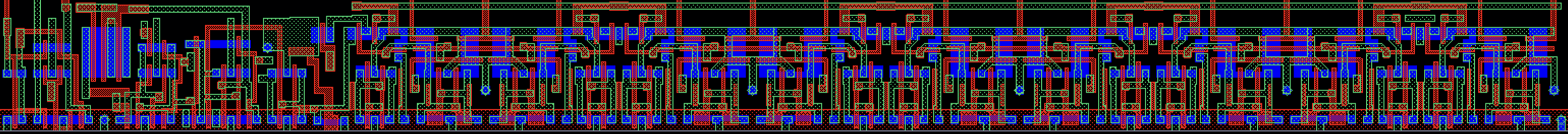}{fig:pixellayout}{Layout of the readout of 1/4 of the group of 4 pixels shown in \fig{fig:pixelgroup}, rotated by 90 degree. Only transistors, poly and metal 1 are shown. The ratio of NMOS:PMOS devices has been designed to be roughly 2:1 so that all devices can be placed in three rows, leading to a very dense layout. Metal 2 is used for local interconnect, while metals 3 and 4 connect signals and power to the bottom of the chip.}{1.0}

\FigureVariable{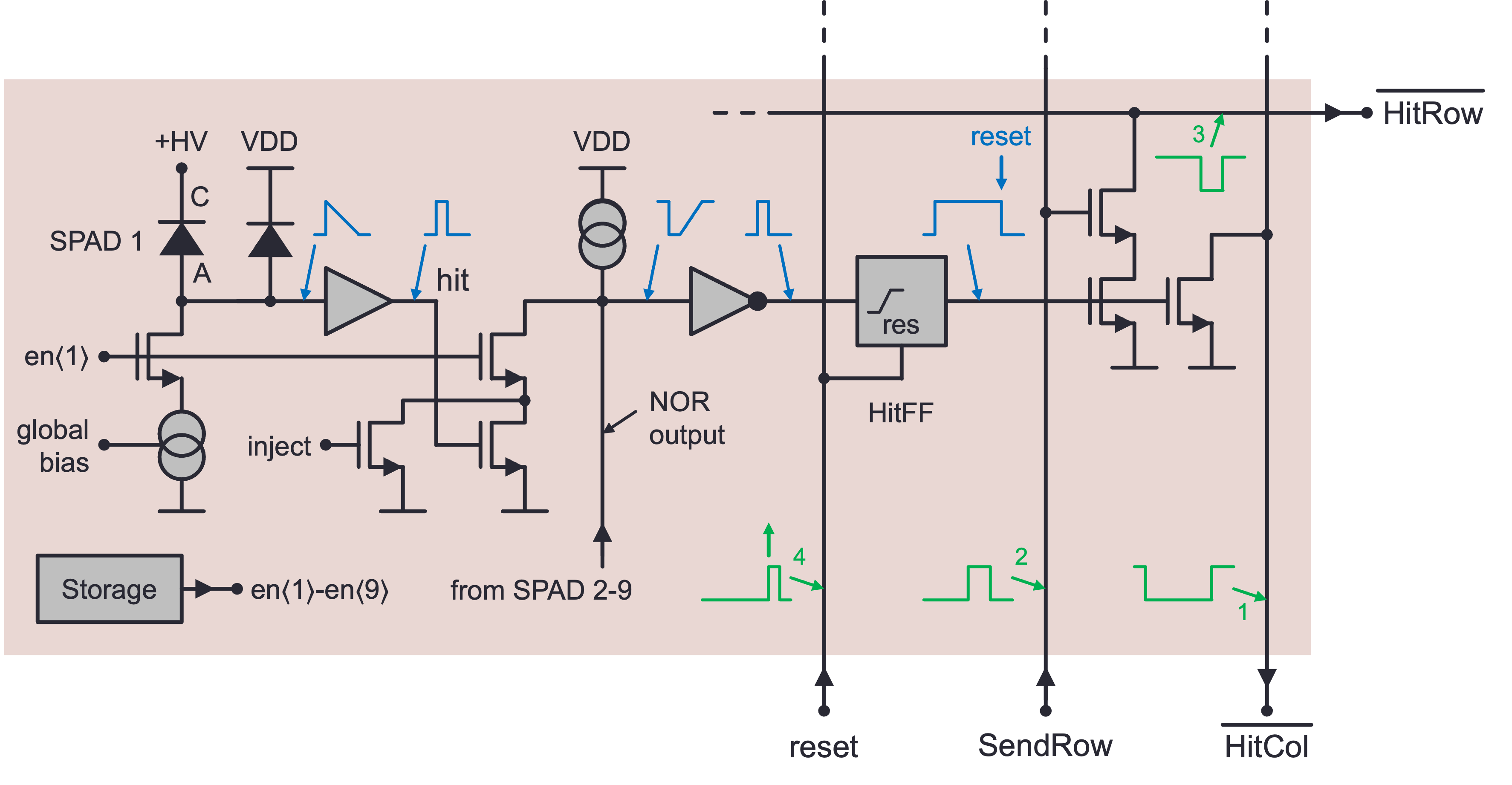}{fig:pixelcircuit}{Circuit in one pixel. For a description, see text.}{0.95}

%---------------------------------------------------------------------------
\subsection{Pixel Circuit}
%---------------------------------------------------------------------------

\newcommand\en[1]{\texttt{en}$\langle$\texttt{#1}$\rangle$}

% Anode = p+, Anode positiv -> Durchlass

The simplified circuit diagram of one pixel (4 of which are contained in the central part of \fig{fig:pixelgroup}) is shown in \fig{fig:pixelcircuit}. The anode of each SPAD is normally pulled to ground with a current source (bottom left, the current can be set globally by an on-chip 7 bit DAC) while the cathode (the N-well common to multiple SPADs) is connected to an external positive bias voltage of $HV\!\approx25\V$. A discharge of the SPAD leads to a large positive signal (equal to the overvoltage) which can be detected by a simple digital buffer consisting of two inverters. No amplifier is required! This hit signal drives a pulldown NMOS which, in combination with a (fixed) PMOS current source, composes a 9-input NOR-gate, merging the hit signals of the $3\times3$ SPADs. A SPAD can be disabled by disconnecting its anode from the current source with a NMOS switch transistor, controlled by one of 9 enable signals \en{1}\dots\en{9}. The anode of the disabled SPAD floats to a positive voltage due to leakage, thus reducing its overvoltage and disabling the gain. A clamping diode assures that the input voltage does not drift too high such that gates could be destroyed. Because a disabled SPAD would generate a hit, the pulldown NMOS of the NOR gate is disabled by the same \en{i} signal.

It is possible to electronically inject a digital hit, independent of SPAD operation, by bypassing the pulldown NMOS of one of the SPADs. By enabling/disabling that particular SPAD, an arbitrary injection pattern can be generated for chip testing.

The output of the wired-NOR gate in connected to the clock input of an edge-triggered hit flipflop (HitFF). A hit in any of the enabled SPADs sets this flipflop to 1 to indicate that a hit has occurred in the pixel. The HitFF will only be cleared once the hit has been read out, so that no hit can get lost. As soon as the HitFF is set, an NMOS pulls down the \HITCOL signal which runs {\em vertically} across the full pixel matrix, as indicated in \fig{fig:pixelcircuit} (right). This can happen in several rows of a particular column. The logic at the chip bottom therefore only knows that there has been $\geq 1$ hit in that column, but it does not know how many, and in which rows. Multiple columns can flag hits. The global logic, described in more detail below, therefore picks {\em one} of the active columns and issues the \SENDROW signal {\em to that particular column}. This signal enables a second pulldown NMOS in each pixel of that column and all pixels drive their HitFF status to {\em horizontally} running \HITROW signals. The logic described in the next section studies that pattern and thereby knows which pixels in the selected column are hit. After recording all hits in the active column, the HitFFs are cleared by issuing the \RESET signal {\em in that column}. The \SENDROW signal is released and the next column with hits is processed.

The 9 control bits required to enable/disable the SPADs are stored in 9 compact static registers also contained in the layout in \fig{fig:pixellayout}. Writing and reading of these registers uses 5 slow strobe signals to select the bit and combinations of the (vertical) \RESET and the (horizontal) \INJECT signal to x/y-address the pixel. This double usage of signals reduces the routing effort.

The circuit for one pixel, as roughly shown in \fig{fig:pixellayout}, contains 199 transistors out of which $45\%$ are used for the 9 masking bits and $36\%$ for the logic per SPAD. 

Note that this circuit does {\em not consume any static power} as long as no hit is present. 

\FigureStandard{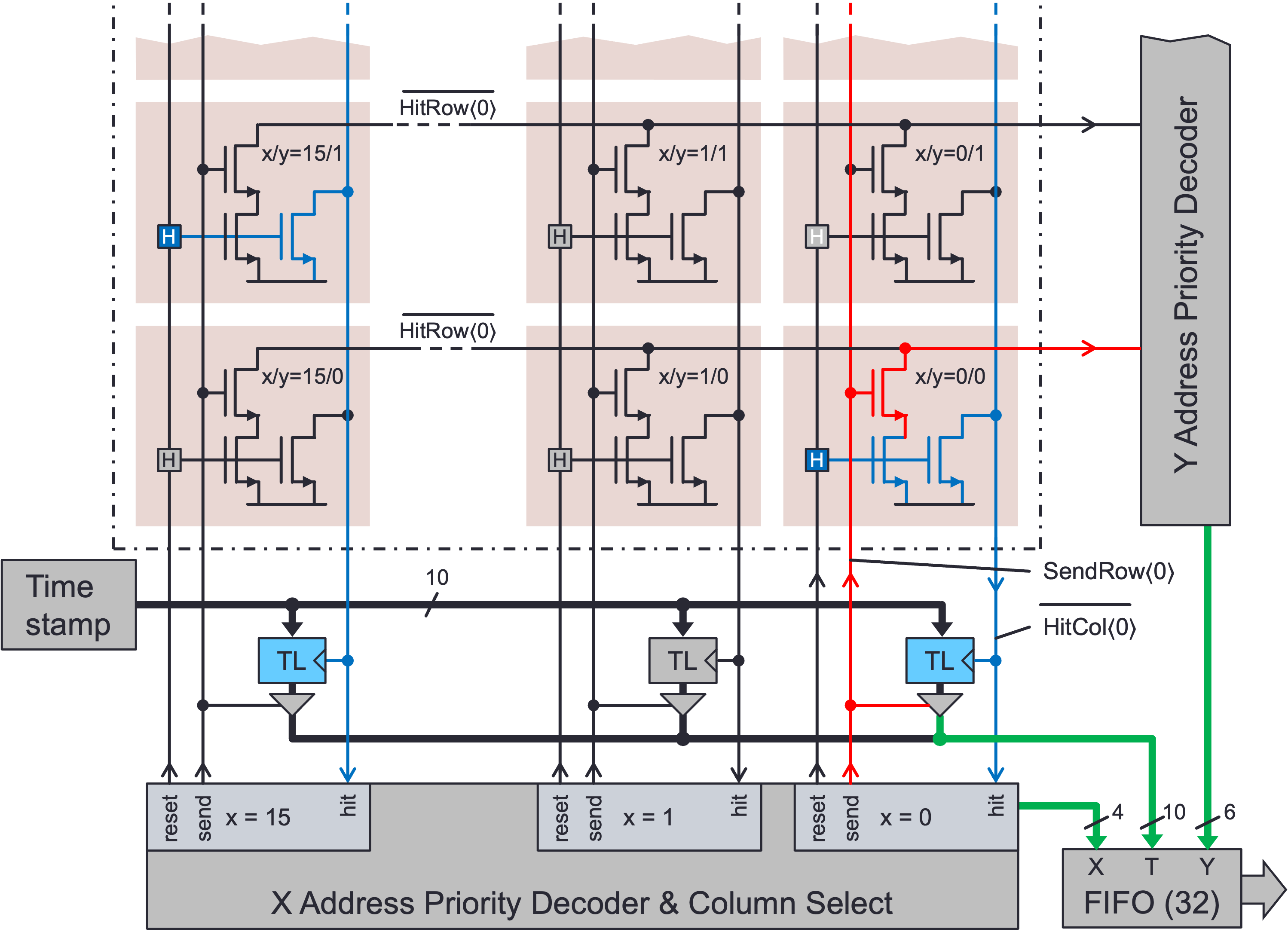}{fig:matrixreadout}{Simplified block diagram of the matrix readout. As an example, the two pixels at $x/y=0/0$ and $15/1$ have been hit. They have communicated this to the x-decoder on the two (blue) vertical \HITCOL signals and the timestamps of the hits have been stored. The x-decoder has selected column 0 for readout by issuing the (red) \SENDROW signal. The corresponding x-address and the stored timestamp are sent to the FIFO (green). The hit pixels {\em in the selected column} activate the (red) \HITROW signals. The y-decoder successively sends the y-addresses of the hit rows to the FIFO (green).}

%---------------------------------------------------------------------------
\subsection{Matrix Readout}
%---------------------------------------------------------------------------

The matrix is assembled from $8\times 30$ groups of $4\times 1$ pixels as shown in \fig{fig:pixelgroup}. For routing reasons, each group has only two vertical \HITCOL signals, but two horizontal \HITROW signals, so that the physical $4\times 1$ pixels appear as $2\times 2$ 'logical' pixels. The encoding of the 16 resulting \HITCOL signals delivers $4$ logical 'x-address' bits, the 60 \HITROW signals require $6$ logical 'y-address' bits. These 10 address bits can be mapped offline to the physical x/y coordinate of the pixel. 

Hit information is communicated from the matrix to the global readout by means of the 
\HITCOL and \HITROW signals, which are pulled up to \VDD by PMOS transistors with a programmable current (7 bit on-chip DAC). When pixels pull the signals towards ground, this bias current flows until the pixels are cleared. This leads to a rate-dependent contribution to the power dissipation. 

The logic to read out the matrix is shown in \fig{fig:matrixreadout}.
A global time information of 10 bit width is fed to 16 time latches (TL in \fig{fig:matrixreadout}), {\em one} for each {\em logical} column. In one operation mode, a 10 bit wide Gray counter is clocked with the system clock of $\approx 50\MHz$ so that the time resolution is $\approx 20\ns$. In a second mode, 9 Gray bits are used and the LSB is provided by the clock itself, so that the time resolution is improved to $\approx 10\ns$.

As soon a hit occurs, the \HITCOL signal in the corresponding column is activated and the hit arrival time is stored in the corresponding time latch. This time information will be associated to {\em all} hits read out from that column during one readout cycle. As a consequence, further hits in a column which arrive between the first hit and the time of readout are associated with a wrong (too early) time stamp.

The readout is controlled by a main finite state machine (MFSM) which 'searches' for hits in the matrix, determines their x/y addresses and time stamps and stores them in a 32 words deep FIFO memory. The MFSM cycles through the following (simplified) states:

\def\WAITHIT  {\texttt{WAIT\_FOR\_HIT}\xspace}
\def\FREEZECOL{\texttt{FREEZE\_COLPATTERN}\xspace}
\def\NEXTCOL  {\texttt{NEXT\_COLUMN}\xspace}
\def\FREEZEROW{\texttt{FREEZE\_ROWPATTERN}\xspace}
\def\SCANROW  {\texttt{SCAN\_ROW}\xspace}
\def\CLEARCOL {\texttt{CLEAR\_COL}\xspace}

\begin{itemize}
\item \WAITHIT:
      This is the starting state, taken also after a \texttt{ResetAll} command to the
      chip (see later). 
      When hits occur in the matrix, one or more corresponding vertically running
      \HITCOL signals are activated by the pixel circuitry described above. The edges of these signals freeze the hit timestamps in the corresponding time latches. A large x-decoder OR-es all \HITCOL signals to generate an overall \texttt{hit} signal, which sends the MFSM to state \FREEZECOL.

\item \FREEZECOL:
      The x-decoder freezes the \HITCOL pattern behind its input. Further 
      \HITCOL signals are ignored (they will be processed later) so that the following readout steps are not disturbed.
      A column index is initialized to the first column 0 and the MFSM jumps to \NEXTCOL. 

\item \NEXTCOL:
      The decoder searches for the next active \HITCOL signal in the {\em frozen}
      input pattern, staring at the column index.
      \begin{itemize}
      \item If no further active column is found, the matrix search is complete and the MFSM jumps to \WAITHIT. If further hits have occurred while the frozen pattern has been processed, these will then immediately be processed by jumping again to \FREEZECOL.
      \item If an active column is found, the column index is set to that column. The scanner presents the corresponding x-address and the associated time latch value to the FIFO. 
      It is known so far that this column has one or more hits, but not yet in which row(s). The decoder therefore issues the \SENDROW signal up the {\em active} column and the MFSM jumps to \FREEZEROW.
      \end{itemize}

\item \FREEZEROW:
      In reaction to the \SENDROW signal in the active column (red signal in \fig{fig:matrixreadout}), the hit pixels have activated their horizontal \HITROW signals. The resulting 60 bit wide pattern is frozen at the input of the 'y-address priority decoder' and the MFSM jumps to state \texttt{SCAN\_ROW}.
      
\item \SCANROW:
      The y decoder finds the first active row in the {\em frozen} input pattern and outputs the corresponding y-address to the FIFO. The FIFO input information (x,y,t) for that hit is now complete and is written to the FIFO. If no space is left in the FIFO, the readout pauses here. The y decoder then clears the just processed row from the frozen input pattern. If there are further active rows, the MFSM remains in state \SCANROW and repeats the described steps, writing successively all hits of the active column to the FIFO. When all hits are processed, the MFSM jump to state \CLEARCOL.

\item \CLEARCOL:
      All hits in the pixels of the active column are cleared by issuing the corresponding column \RESET signal.
      The time latch of that column is cleared, and the hit flag in the frozen column pattern is cleared.
      The MFSM then jumps to state \NEXTCOL.

\end{itemize}

This readout guaranties that every set hit flipflop is read out. The complete readout cycle described above requires 7 clock cycles for a single hit, so that roughly 7 million hits per second can be extracted from the matrix at a $50\MHz$ clock speed. These hits are buffered in the FIFO, which is emptied at a slower rate (see later) so that the actual maximum hit rate is lower. Multiple hits in close time sequence are processed with correct time stamps, if they occur in separate columns, and if there is no general data congestion. As mentioned above, a high number of more or less simultaneous hits in one column (S2 signals) will lead to wrong time stamps for later hits. Note that the presence of multiple hits in one column in one readout cycle is known from the data so that timestamp issues are flagged. 
The impact of these effects at high rate conditions on the data analysis remains to be studied.
 
\FigureVariable{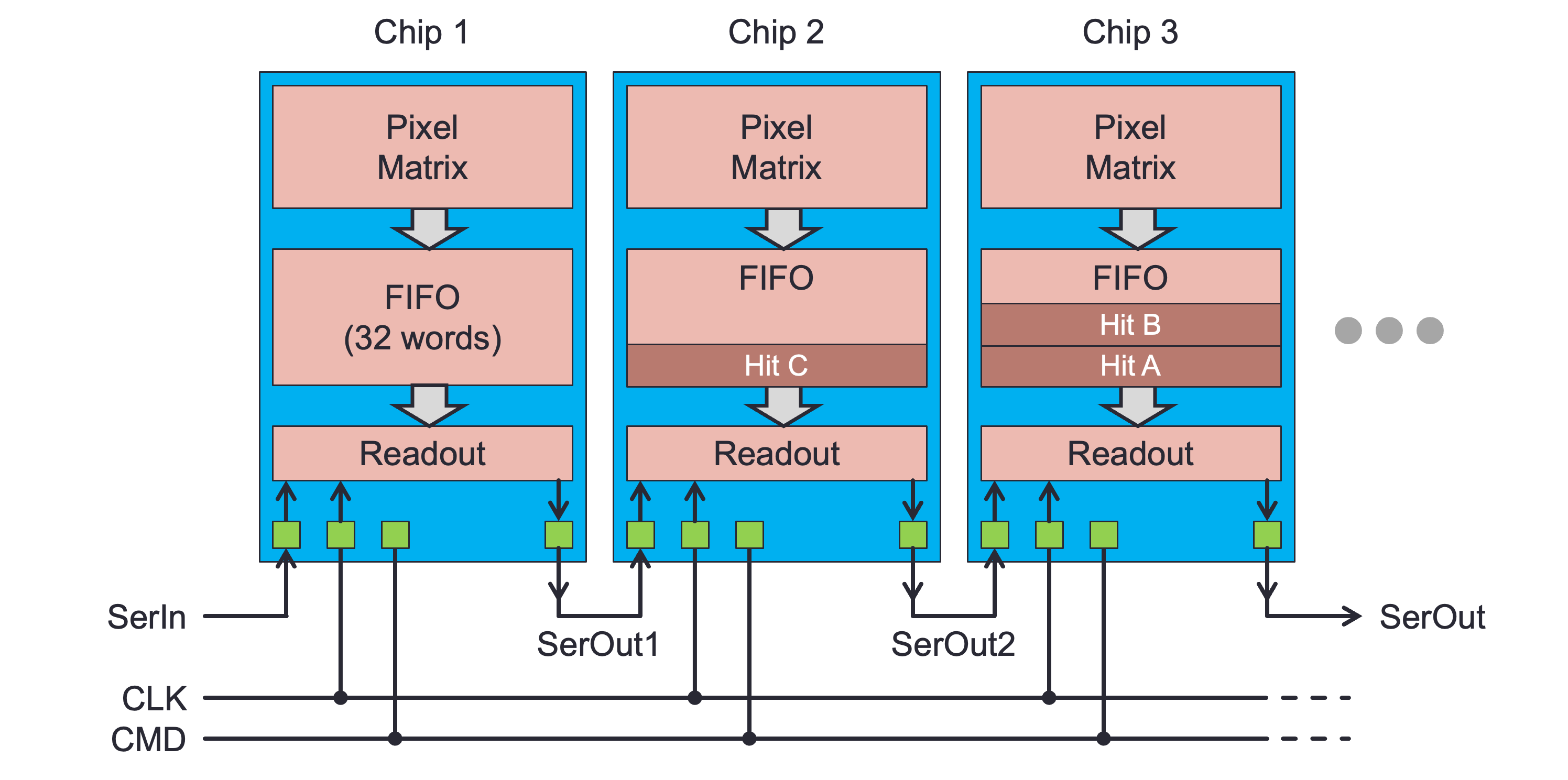}{fig:chipchain}{Many chips can be connected in series by sending the \SEROUT signal of one chip to the \SERIN input of another chip. The clock \CLK and the command signal \CMD are shared by all chips. In this example, the chips have 0,1,2 hits in their FIFOs waiting for readout.}{0.85}

%---------------------------------------------------------------------------
\subsection{Hit Data Readout}
%---------------------------------------------------------------------------

In order to simplify the on-chip logic as much as possible, and to minimize the number of chip pins and signals, a custom protocol has been implemented for chip readout. All chips receive the same \CLK signal of $50\MHz$ or more. Hit data are sent out serially on pin \SEROUT. In order to allow for chip daisy-chaining, this serial bit stream can be injected into the \SERIN input of another chip, which then 'merges' it with its own hit data and passes everything to its output. In this way, many chips can be connected in series, as illustrated in \fig{fig:chipchain}.

\begin{figure}[!t]
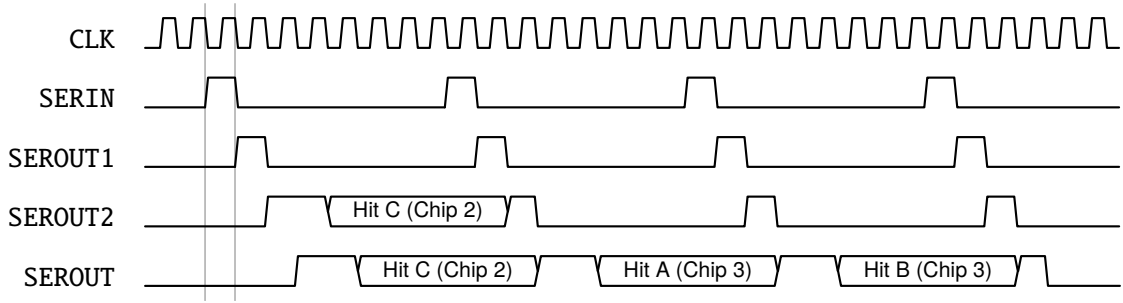

\centering
\begin{tikztimingtable}[scale=1.3]  % NOTE: Empty lines are NOT allowed in this block!!!
  \CLK    & l 64{t}\\
  \SERIN  & 2L     {H7L}                H7L                  H7L                  H5Ll\\
  \SOA    & 2L L   {H7L}                H7L                  H7L                  H4Ll\\
  \SOB    & 2L LL  HH6D{[scale=1.3]Hit C (Chip 2)} H7L                  H7L                  H3Ll\\
  \SEROUT & 2L LLL HH6D{[scale=1.3]Hit C (Chip 2)} HH6D{[scale=1.3]Hit A (Chip 3)} HH6D{[scale=1.3]Hit B (Chip 3)} H2Ll\\
  \extracode
  %\draw (0,0) circle (0.9pt); % show origin
  \begin{pgfonlayer}{background}
  \vertlines[help lines]{2,3}
  \end{pgfonlayer}
\end{tikztimingtable}
\caption{Hit readout: The first chip in the chain receives regular isolated \logicone signals
indicating the start of an empty data packet (the length of the packets is reduced in the figure for illustration purpose). In the example (see also \fig{fig:chipchain}), chip 1 has no hits and passes the pattern unchanged. Chip 2 has a hit and injects it in the first empty packet by adding a second \logicone level and the payload. Because chip 2 has not further hits in its FIFO, it passes the subsequent empty packets unchanged. Chip 3 has 2 hits and uses the next two packets for sending.}
\label{fig:readout}
\end{figure}

Hit data are transported in \inquotes{packets} of a fixed size of $28$ bit. The start of a data packet is indicated by a logic \logicone at \SERIN. The next bit in the serial data stream determines whether the data packet is empty or contains valid hit data:
\begin{itemize}
\item A \logicone level flags a full data packet.
      The two bits are then followed by 26 bit of payload data:
      10 bit for the x/y hit position, 10 bit for the time stamp, and 6 bit for a chip ID.
      When the readout logic of a chip detects an incoming full packet, it passes all 28 bits untouched to the output.
\item A \logiczero level flags an empty packet.
      If the chip has hit data in its FIFO, it flips the bit to \logicone and then injects the 26 bits of hit data into the serial data stream, so that a new 'full' packet is generated. If the FIFO is empty, the empty packet is passed unchanged to the next chip.
\end{itemize}
This mechanism is illustrated in \fig{fig:readout}. Note that the first chip must receive a 
\logicone level every 28 clocks at its \SERIN input to define the data packet structure.

At the anticipated clock speed of $50\MHz$, the readout of one hit requires $28\times20\ns = 560\ns$ so that at most $1.8\Mhits/\s$ can be handled by one chain of serially connected chips. The readout mechanism presented here leads to an unequal emptying of the chips in a chain: The first chip receives only empty packets, so that it can empty its FIFO at the maximum rate, while the last chip in the chain sees many full packets and has to wait longer for empty packets. Thanks to the hit buffering in the FIFOs, this effect will only be significant when the system reaches its bandwidth limit. A fair assignment of empty packets to the chips can be obtained if the chips at the start of the chain are forced to forward a certain number of empty packets downstream even if their FIFOs have data \cite{FDCS}. Such a mechanism may be implemented in later chips.

%---------------------------------------------------------------------------
\subsection{Chip Control}
\begin{figure}[!t]
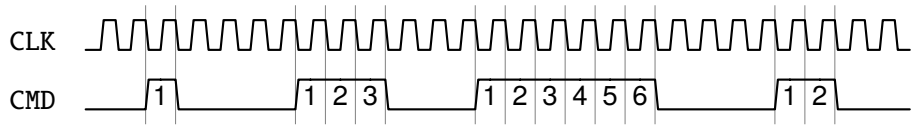

\centering
\begin{tikztimingtable} [scale=1.3,
  % NOTE: Empty lines are NOT allowed in this block!!!
% from
% https://tex.stackexchange.com/questions/37254/tikz-timing-text-labels-in-symbols-other-than-d-and-time-axis-discontinuities
  timing/metachar={{K}[2]{#1h !{++(0,-.5\yunit)} N[rectangle,scale=1.0]{#2} !{++(0,+.5\yunit)} #1h}}
  ]
  \CLK    & l54{t}\\
  \CMD    & 2L K{1}  4L K{1}K{2}K{3} 3L K{1}K{2}K{3}K{4}K{5}K{6} 4L K{1}K{2}2Ll\\
  \extracode
  \begin{pgfonlayer}{background}
  \vertlines[help lines]{2,3, 7,8,9,10, 13,14,15,16,17,18,19, 23,24,25}
  \end{pgfonlayer}
\end{tikztimingtable}
 \caption{The \CMD signal is sent to all chips in parallel. It is normally 0. When going high for a number of \CLK cycles, it triggers a particular action in all chips in parallel after its falling edge.} \label{fig:control}
\end{figure}

%---------------------------------------------------------------------------
%
Chip operation and testing would require a number of further signals, for instance to reset the time counters or to perform test injections. In addition, the mask pattern for the SPADs must be written individually to each chip. In order to save signals and wire bonds, a custom protocol for chip control has been implemented. It uses a global \CMD signal which is broadcast to all chips synchronous to the global clock \CLK, see \fig{fig:chipchain}. \CMD is normally \logiczero. It is then pulsed to \logicone for a certain number of clock cycles. All chips count how long the signal was active and take different actions as a function of \CMD duration. \Fig{fig:control} illustrates this pulse-width encoding scheme.
	
\begin{table}[!t]
\centering
\begin{tabular}{llp{0.5\textwidth}} 
\toprule
 \CMD  & Name                      & Action \\
 width &                           &        \\
\midrule
    1  & \texttt{ResetAll        } & Reset State machines, FIFO, time counter and hits\\
    2  & \texttt{ResetTime       } & Reset only time counter\\
    3  & \texttt{ResetMatrix     } & Reset only hits in matrix\\
    4  & \texttt{ReadoutSimple   } & Start readout of only one hit\\
    5  & \texttt{StartReadout    } & Start continuous hit readout\\
    6  & \texttt{StopReadout     } & Stop  continuous hit readout\\
    7  & \texttt{WriteConfig     } & Write configuration register\\
    8  & \texttt{ReadConfig      } & Read  configuration register\\
    9  & \texttt{WriteID         } & Write chip IDs\\
%All Chips wait for ID. They take input data, but keep output at 0 as long as they have no receivede and ID. user injects 1st ID into chain -> first chip takes ID and leaves 'read ID mode'. User injects next...\\
   10  & \texttt{InjectMatrix    } & Inject hits into the matrix, depending on the programmed 
                                     enable pattern\\
   11  & \texttt{InjectFIFO      } & Inject test data pattern to FIFO\\
   12  & \texttt{InjectSerializer} & Inject test data pattern to serializer (behind FIFO)
\\
\bottomrule
\end{tabular}
\caption{The 12 available commands are pulse-length coded by the \CMD signal.}
\label{table:commands}
\end{table}

The chip presently implements 12 different commands which are listed and described in \tab{table:commands}. Most commands just initiate one particular action (reset, hit injection, test pattern generation) in all chips in parallel. Some commands are more complicated, because they act on individual chips. In these cases, the existing serial data chain is used to send or receive not-hit-data to/from the chips. Three important commands are described here in more detail:

\begin{itemize}
\item \texttt{WriteID}: This command is used to program the ID of each chip.
      When it is detected, {\em all} chips go to a special 'wait-for-ID' state.
      Each chip then monitors its serial input. When it detects an empty data packet, it takes the content,
      stores some of the bits in its ID register, passes it as {\em full} packet to the downstream chips, and leaves the 'wait-for-ID' state. All subsequent packets are passed downstream. In this way, all chip IDs can be set externally. The first ID sent will be consumed by the first chip in the chain, the next one by the second chip and so on. This method to fix chips IDs has a very wide addressing range and does not need any additional pads.
\item \texttt{WriteConfig}:
      This command can only be used after each chips has received a unique ID.
      When \texttt{WriteConfig} is issued, all chips go to a 'write-config' state and continuously monitor the serial data stream. When a valid packet is observed, a target ID coded in the payload is compared to the chip ID. When both match, the remaining bits (containing internal addresses and data) are used to write chip registers, like for instance the two 7 bit DACs (SPAD recharging current and bus pullup current) or the SPAD masking bits. The 'write-config' state is only left when another command is issued, so that more addresses or chips can be written.
\item \texttt{ReadConfig}:
      In order to verify that the configuration was successful, all   
      global configuration bits can be read back. (The masking bits in the matrix cannot be read.) The mechanism is similar as for writing: After the
      chip has been set to the 'read-config' state, it waits for data packets containing its ID. If such a packet is detected, the address contained in the package is used to retrieve register data and the corresponding value is written to the data section of the packet.
\end{itemize}

%===========================================================================
\section{First Chip Test Results} \label{sect:measurements} 
%===========================================================================

% 22.1.2025: DCR on PCB: 40kHz/mm² (gar nicht so weit weg von den 25kHz/mm²)

\FigureVariable{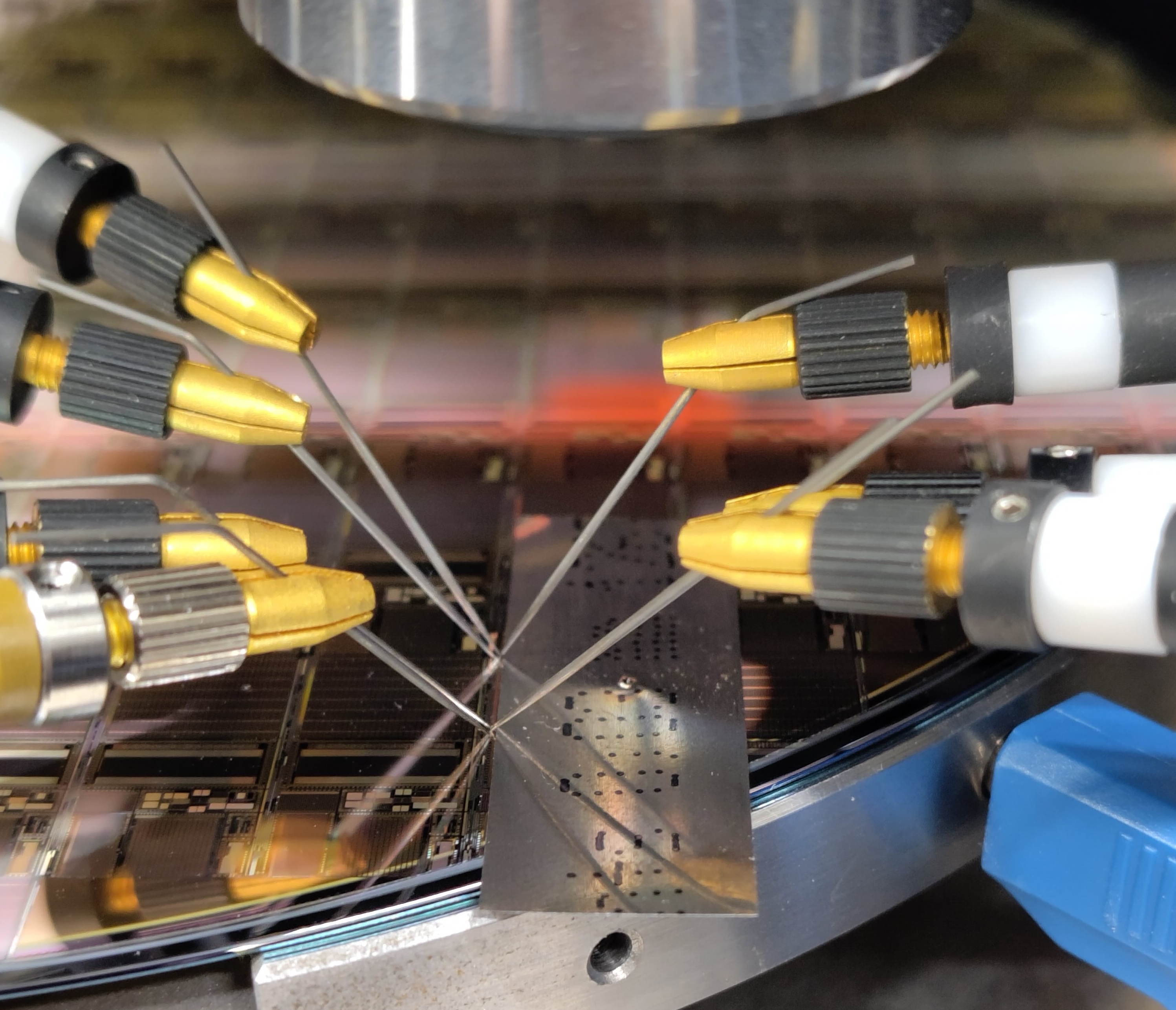}{fig:prober}{Initial chip tests were done on a wafer prober using 7 needles. A metal mask with a small hole was placed on the chip/wafer to demonstrate position sensitive photon detection.}{0.6}

The engineering run was completed shortly before the conference, so that no diced chips were available yet. The first functional tests were therefore done on a wafer prober, requiring only 7 needles thanks to the simple digital protocol, see \fig{fig:prober}. The signal- and power-integrity in this setup is poor, so that a low clock speeds of a few $\MHz$ has been used.

\begin{figure}[!t]
 \centering
 \includegraphics[width=0.45\linewidth, trim = 0 -50 0 0]{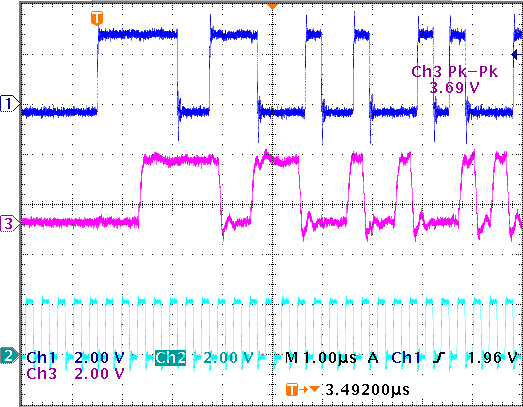}
 \hfill
 \includegraphics[width=0.50\linewidth]{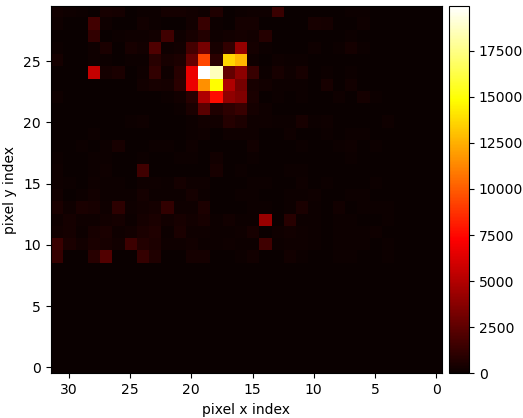}
 \vskip-2mm
 \caption{Left: Transmission of a serial digital pattern from \SERIN (top trace, blue) to \SEROUT (middle trace, magenta) at slow \CLK (bottom trace, cyan) speed. Signal integrity is poor due to the prober setup. Right: Acquisition of SPAD hits. The increased dark hit rate under the hole in the metal mask is clearly visible. The SPADs at the bottom close to the IO pads (needles!) were masked because they were not sufficiently light-shielded.}   
 \label{fig:results}
\end{figure}

The digital control part works as expected. As an example, \fig{fig:results} (left) shows the transfer of a bit pattern though the chip from \SERIN to \SEROUT. A chip ID can be written and control registers can be written and read back. Correct operation of the SPADs and of the matrix readout was confirmed by illuminating a small fraction of the chip through a metal mask placed directly on the wafer, see \fig{fig:prober} and \fig{fig:results} (right).

%===========================================================================
\section{Discussion and Outlook} \label{sect:discussion} 
%===========================================================================

The Digital SiPM technology of the Fraunhofer Institute IMS (Duisburg, Germany) offers SPADs optimized for cold operation with a dark count rate of only $\approx 0.02 \Hz$ per $\mm^2$ of active area at liquid Xenon temperature and of $\approx 0.01 \Hz / \mathrm{mm}^2$ at liquid Argon temperature. The crosstalk between adjacent SPADs on a test chip is $\lesssim 4\%$ (paper in preparation). 

% Normaler Prozess: 0.4Hz/mm2 (@ LXe), 0.05Hz/mm2 (@ LAr)

The presented chip contains pixels with an average size of $250\times291\um^2$ and has an overall fill factor of $\approx 72\%$. Each photon hit is individually read out with that spatial resolution and a time stamp with $\approx 10\ns$ resolution.
The readout is purely digital using only two global signals (\CLK, \CMD) and a serial input/output pair (\SERIN, \SEROUT). Many chips can be connected to one long readout chain so that the number of cables is strongly reduced. The number of chips that can be grouped in one chain is mainly determined by the required data throughput: At a typical clock rate of $50\MHz$, the readout of one photon hit requires $560\ns$, so that a readout chain saturates at $1.8\Mhits/\mathrm{s}$. In our target application, the sparsity of the physical events leads to a low rate of S1 events with few photons each. The overall hit rate is probably dominated by the dark count rate. At a DCR of $0.1 \Hz / \mathrm{mm}^2$, one {\em square meter} of SPADs would generate $100\khits$ per second, which is still far below the bandwidth limit of a {\em single} serial link. This very coarse estimation shows that a very small number of links should be sufficient for handling S1 signals and DCR.

The second type of events in the application are bursts of several 1000 of more or less simultaneous photons in a rather small area of the detector (S2 signal). For {\em very high} local photon densities, the rather large pixel size could lead to some pileup losses. The architecture ensures that all hits are recorded and read out, but some time stamps may by corrupt at very high occupancy. The readout of these many photons takes a significant time ($\approx 0.5\ms$ for 1000 hits) during which (only) the pixels waiting for readout are blind for new hits. The impact of such inefficiencies needs to be studied by simulations. If it turns out to be an issue for the applications, a double hit buffering in the pixels could be introduced with little impact on fill factor (nearly half of the circuit area is presently needed for masking of noisy SPADs).

%---------------------------------------------------------------------------
\subsection{Module Design}
%---------------------------------------------------------------------------

In order to cover larger areas, several chips must be grouped together on \inquotes{modules}. Each such module requires only 7 electrical signals: \GND, \VDD to supply the chip, \BIAS for the SPADs, and \CLK, \CMD, \SERIN and \SEROUT for digital control and data readout. Several modules can again be daisy-chained with no increase in the number of signals as long as the readout bandwidth is sufficient. 

\FigureStandard{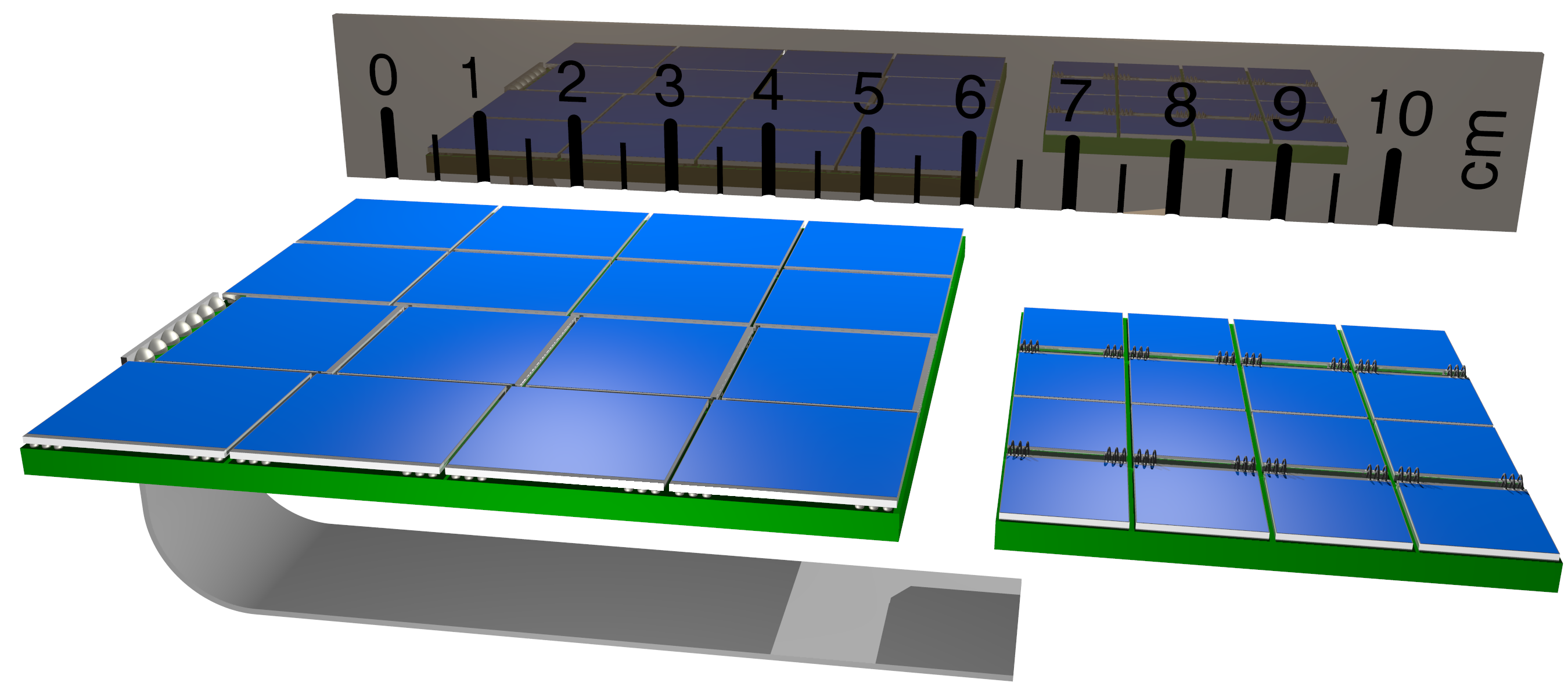}{fig:module}{Concepts for larger Digital SiPM modules. Right: Using the chip presented in this paper. Left: Using a larger chip with through-silicon-vias (TSVs) and bump connections. The gaps between the chips have been exaggerated for better visibility.}

\Fig{fig:module} shows 3D renderings of possible module designs. The right module consists of 16 chips as described in this paper. Pairs of chips are arranged with the wire-bonding side facing each other. The wire bonding connections from the chips to the carrier substrate pass through a gap between the chips of $\approx 0.6\mm$ width. $2\times 4$ of these chip pairs are arranged with a (convenient) chip-chip gap of $100\um$. This leads to a module of $\approx 3.3 \times 3.5 \cm^2$ size with a SPAD fill factor of $\approx \ffModuleA \%$.

A more advanced design is shown on the left side of \fig{fig:module}, where a chip size of $\approx 15 \times 14.6 \mm^2$ is assumed, compatible with the reticle size limitations of the manufacturer (even larger chips would be possible, but production yield may then become an issue). The chips use through-silicon vias (TSVs) to bring the IO signals and power from the front to the backside where they are connected to the substrate with solder balls.
One chip row has been rotated by 90 degrees so that a small region on the substrate is left empty. This area can be used to connect a vertically running flat (caption-type) cable with solder connections. This concept would allow placing modules close to each other with nearly no gap, and it would avoid electrical feed-through connections from the chip side to the backside, so that the substrate only needs to be processed on the front side. This simplifies substrate manufacturing considerably (for instance compared to \cite{SiInterposer}) and opens up a wide choice of materials with low intrinsic radioactivity and a coefficient of thermal expansion (CTE) matched to the chips, an obvious candidate being silicon. Only one routing layer may be required on the substrate if the IO signals are re-arranged with a redistribution layer on the bottom side of the chips. The expected SPAD fill factor of a plane covered by such modules, assuming lateral gaps of $0.5\mm$, is $\approx \ffPlaneB \%$. This is better than what is typically achieved with PMTs\footnote{The popular 3-inch PMT R11410–10 from Hamamatsu has an outer diameter of $76\mm$ and an active area diameter of $65\mm$ \cite{PMT2} so that the best hexagonal packing reaches $\le 64.3\%$.}.

% Sending hit to FIFO takes 160 ns.

% Talk https://indico.cern.ch/event/1458618/  16:45
% Dark count rate PMT   1 Hz /cm2
% FF 62 (round, 3 inch) - 75% (square)

%%%%%%%%%%%%%%%%%%%%%%%%%%%%%%%%%%%%%%
% IRRELEVANT

% Absolute PDE measurement in Nagoya!

% DARWIN Meeting: 73 Leute
% Alessandro Razetto (FBK, INFN, ...)
% - comment: BSI SPADs have 10x higher DCR.
% - comment: EPI must be 15um at least, fully depleted
% THEY: Low DCR rate SiPM with FbK. FBK terribly late... 2025
% THEY: 2.5D SiPM: Commercial SiPM. Active Silicon Carrier. 
% Can disable pixels (not SPADs) FBK slow. Also working with Hamamatsu

% Make better time stamps.

\bibliography{Paper_Ref} 

\end{document}